\documentclass[%
 reprint,
 amsmath,amssymb,
 aps,
prb,
floatfix,
]{revtex4-2}

\usepackage{bm}
\usepackage{braket}
\usepackage{dsfont}
\usepackage{graphicx}
\usepackage[dvipsnames]{xcolor}
\usepackage{color}
\usepackage{comment}
\usepackage{appendix}

\begin{document}

\preprint{APS/123-QED}
\title{Orbital Frustration and Emergent Flat Bands}
 \author{Wenjuan Zhang}%
\thanks{These authors contributed equally}
\affiliation{Department of Physics,
Ohio State University,
Columbus, OH 43210, USA}
\author{Zachariah Addison}%
    \thanks{These authors contributed equally}
\affiliation{Department of Physics,
Ohio State University,
Columbus, OH 43210, USA}
\author{Nandini Trivedi}
\affiliation{Department of Physics,
Ohio State University,
Columbus, OH 43210, USA}

\date{\today}

\begin{abstract}

We expand the concept of frustration in Mott insulators and quantum spin liquids to metals with flat bands. We show that when {\em inter}-orbital hopping $t_2$ dominates over intra-orbital hopping $t_1$, in a multiband system with strong spin-orbit coupling $\lambda$, electronic states with a narrow bandwidth $W\sim t_2^2/\lambda$ are formed compared to a bandwidth of order $t_1$ for intra-orbital hopping. We demonstrate the evolution of the electronic structure, Berry phase distributions for time-reversal and inversion breaking cases, and their imprint on the optical absorption, in a tight binding model of $d$-orbital hopping on a honeycomb lattice.  Going beyond quantum Hall effect and twisted bilayer graphene, we provide an alternative mechanism and a richer materials platform for achieving flat bands poised at the brink of instabilities toward novel correlated and fractionalized metallic phases.  
\end{abstract}

\maketitle

\noindent {\em{Introduction:}} Symmetry and topology have played a fundamental role in characterizing emergent phases in quantum matter. From the well understood emergence of long-range order in crystals, magnets and superconductors, research has moved to the frontiers of emergence of a new type of truly exotic order, topological order in which the quantum numbers of the excitations are fractionalized. A key ingredient to obtain fractionalization is frustration of the motion of the electron that leads to a small kinetic energy and hence to flat bands in the electronic spectrum. Our main result is to provide a design principle for creating materials in which the inter-orbital hopping $t_2$ dominates over intra-orbital hopping $t_1$, in a multiband system with strong spin-orbit coupling $\lambda$, that generates electronic states with a narrow bandwidth $W\sim t_2^2/\lambda$ compared to a bandwidth of order $t_1$ for intra-orbital hopping. We expect these designer flat-band materials to provide a  link between frustrated metals and frustrated magnetic insulators opening a doorway for the search and discovery of topologically ordered states. 

Emergent phases in quantum matter are understood today under two paradigms: the Landau paradigm that has been the pillar for phases arising from the spontaneous breaking of symmetry and the topological paradigm in which phases are characterized by topological invariants leading to quantized response functions. 
Within the topological paradigm there are broadly two classes: phases where interactions are less important, such as the integer quantum Hall effect, topological insulators \cite{hasan2010colloquium, kane2005z}, Weyl semimetals \cite{yan2017topological,lv2015experimental}, and topological superconductors \cite{qi2011topological,sato2017topological}, and the more exotic class, where interactions dominate along with topology, yielding a new type of order, dubbed topological order \cite{chen2013symmetry,levin2006detecting,wen1990topological}. These include two candidates so far: (1) the fractional quantum Hall effect \cite{stormer1999fractional,jain1990theory,haldane1983fractional,wen1995topological} in a two-dimensional electron gas subjected to a large magnetic field in which charge is fractionalized,  and (2) quantum spin liquids in Mott insulators in which spin is fractionalized \cite{savary2016quantum,zhou2017quantum,kitaev2006anyons}. 

Our broad motivation is to search for materials that have flatbands in the absence of a magnetic field in which strong correlations can generate topologically ordered phases that show fractionalization of charge, fractionalization of spin, and quantized anomalous and topological Hall effects. 

\begin{figure*}[tbh!]
\includegraphics[width = 180mm]{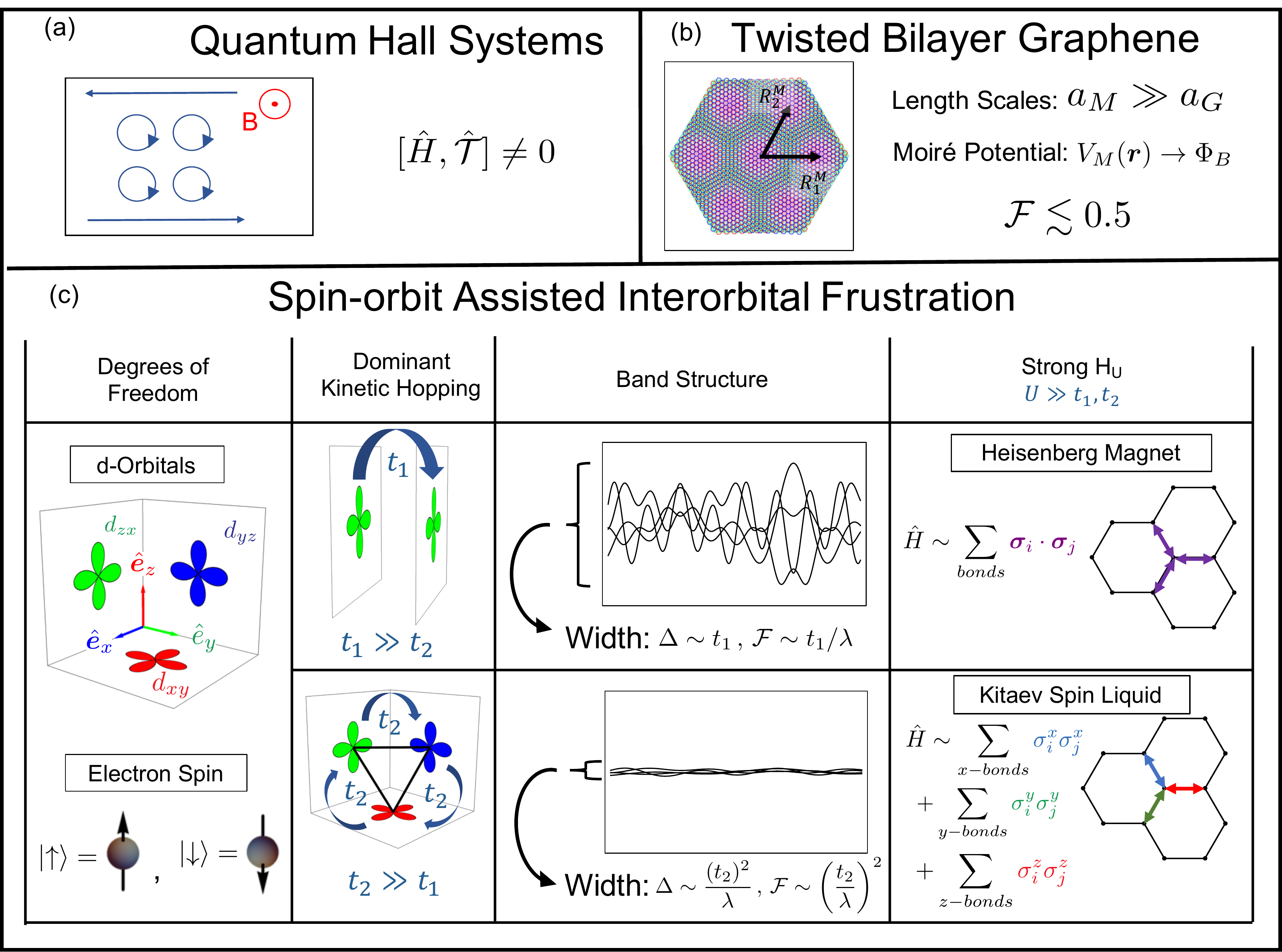}
\caption{\label{flatbands} Figure showing three mechanisms for generating flat bands. (a) In quantum Hall systems flat bands emerge when a large magnetic field $\bm{B}$ is applied to a two-dimensional electron gas that localizes the orbital motion of the electrons in the bulk of the material. (b) Flat bands in twisted bilayer graphene emerge from a periodic interlayer Moiré potential $V_M(\bm{r})$ whose period $a_M$ is much larger than the intralayer graphene lattice constant $a_G$.  When the angle of rotational misalignment between the intralayer lattice vectors is equal to a {\it magic} angle , the potential acts as an effective time preserving pseudo flux $\Phi_B$ that greatly reduces the kinetic energy of the electron \cite{PhysRevLett.122.106405,san2012non,liu2019pseudo}. These systems have a flatness ratio of $\mathcal{F}\lesssim 0.5$.  (c) In our model flat bands are achieved through spin-orbit assisted inter-orbital frustration whereby dominant inter-orbital kinetic hopping can quench the kinetic energy of electrons in an effective $j=1/2$ multiplet via a large onsite spin-orbit interaction on the honeycomb lattice leading to a flatness ratio $\mathcal{F}\sim(t_2/\lambda)^2$ tuned by the ratio of the inter-orbital kinetic hopping strength $t_2$ and the spin orbit interaction $\lambda$.  The strong interacting limit of dominant intra-orbital or dominant inter-orbital hopping leads to an effective Heisenberg Hamiltonian or an effective Kitaev Hamiltonian respectively.}
\end{figure*}

To put this in perspective, fractionalization in FQHE arises because the kinetic energy is quenched and the bands become completely flat in a high magnetic field upon projecting to the lowest Landau level (see Fig.~\ref{flatbands} (a)), and consequently, the relatively strong electron-electron Coulomb interactions open a bulk gap in the metallic regime with edge states that carry the signature of fractional charge in quantized Hall and quantized thermal Hall transport. 

Quantum spin liquids are expected in strongly correlated systems, in which a Mott gap separates the high energy charge dynamics from the low energy spin dynamics. Further, the magnetic interactions are frustrated either from the geometry, or from competing interactions, that can lead to spin textures and quantum spin liquid phases~\cite{ronquillo2019,patel2019,doretto2014plaquette,sachdev1990bond,singh1999dimer,kotov1999low,zhitomirsky1996valence,ralko2009generalized,capriotti2001resonating,jiang2012spin,li2012gapped}. In Kitaev spin liquids on the honeycomb geometry, frustration arises due to bond-dependent interactions arising from strong spin-orbit coupling. As a result of the strong frustration, spins are fractionalized into majorana modes that have recently been shown to produce a quantized thermal Hall conductance in a candidate material $\alpha$--${\rm RuCl}_3$ \cite{kitaev2006anyons,nasu2017thermal,pidatella2019heat,laurell2020dynamical,do2017incarnation}.

One route is to investigate transition metal based materials that have narrow bands and hence strong Coulomb interactions $U$ compared to the wider-band s- and p- band materials. For example, in the cuprate superconductors, the bandwidth is $W=8t \approx 2.5 eV$ and U is ~5 eV giving a ratio $U/W\approx 2$. Tunability of $U$ is difficult as it is determined by the atomic correlations in transition metal compounds, so a complimentary approach is to reduce the bandwidth $W$ in order to enhance the ratio $U/W$. 

Recently, the community has seen feverish activity in twisted bilayer graphene in which flat bands emerge when the angle of mismatch is a very specific {\it magic} angle of around $1^\circ$ yielding a large scale moiré pattern with lattice constant much larger than that of a graphene sheet ~\cite{bistritzer2011moire,dos2012continuum}; (see Fig.~\ref{flatbands} (b)). For bilayers encapsulated by hBN the long ranged Coulomb energy scale is $U\sim 10-15$ meV and the bandwidths of the flattest bands are $W \lesssim 5$ meV \cite{PhysRevB.102.045107}. For the most ideal {\it magic} angles the flat bands are separated from other dispersing bands by a energy gap $\Delta\sim 10$ meV leading to a flatness ratio of $\mathcal{F}=W/\Delta\lesssim 0.5$ \cite{PhysRevB.96.075311,PhysRevB.102.201112}.  The flat bands have resulted in the observation of strongly correlated superconducting and Mott insulating phases~ \cite{cao2018unconventional,cao2018correlated, xu2018topological,ochi2018possible,wu2018theory,zhang2018moire,pizarro2019nature}; however so far fractionalized metallic states have not been observed.

In this paper we demonstrate a new mechanism for achieving flat bands in {\em metallic} systems with strong spin-orbit coupling. The basic mechanism involves frustrating 
the motion of an electron by suppressing intra-orbital hopping compared to hopping between different orbitals; we dub this mechanism {\em spin-orbit assisted inter-orbital frustration} (see Fig.~ \ref{flatbands} (c)). 
We demonstrate the emergence of flatbands in the metallic regime, with the promise of fractionalized metals and other exotic magnetic phases upon including correlations, in a tightbinding model of $d$-orbitals on a honeycomb lattice with strong spin-orbit coupling.

We choose the parameters of our tightbinding Hamiltonian to ensure that with on-site Hubbard interactions, in the strong coupling limit, we obtain the Kitaev model with bond-dependent interactions and contrast it with the parameters that generate the isotropic Heisenberg model (see Appendix \ref{AppC2}). Our main result is for the metallic itinerant phase, where the presence of strong spin-orbit interactions, $\lambda$, can lead to frustration whereby the kinetic energy of a multiplet of bands is suppressed leading to small bandwidths of order $(t_2)^2/\lambda$ and small flatness ratios $\mathcal{F}\sim (t_2/\lambda)^2$ in systems where intersite {\em inter}-orbital hopping, $t_2$, dominates over intersite {\it intra}-orbital hopping $t_1$.  We calculate the electrical conductivity in two regimes: systems dominated by $t_2$ 
and systems dominated by $t_1$,
for which we show the crossover and consequences of the momentum space Berry phase distributions in the emergent flat band phase.

\begin{figure*}[htb!]
    \centering
    \includegraphics[width=\textwidth]{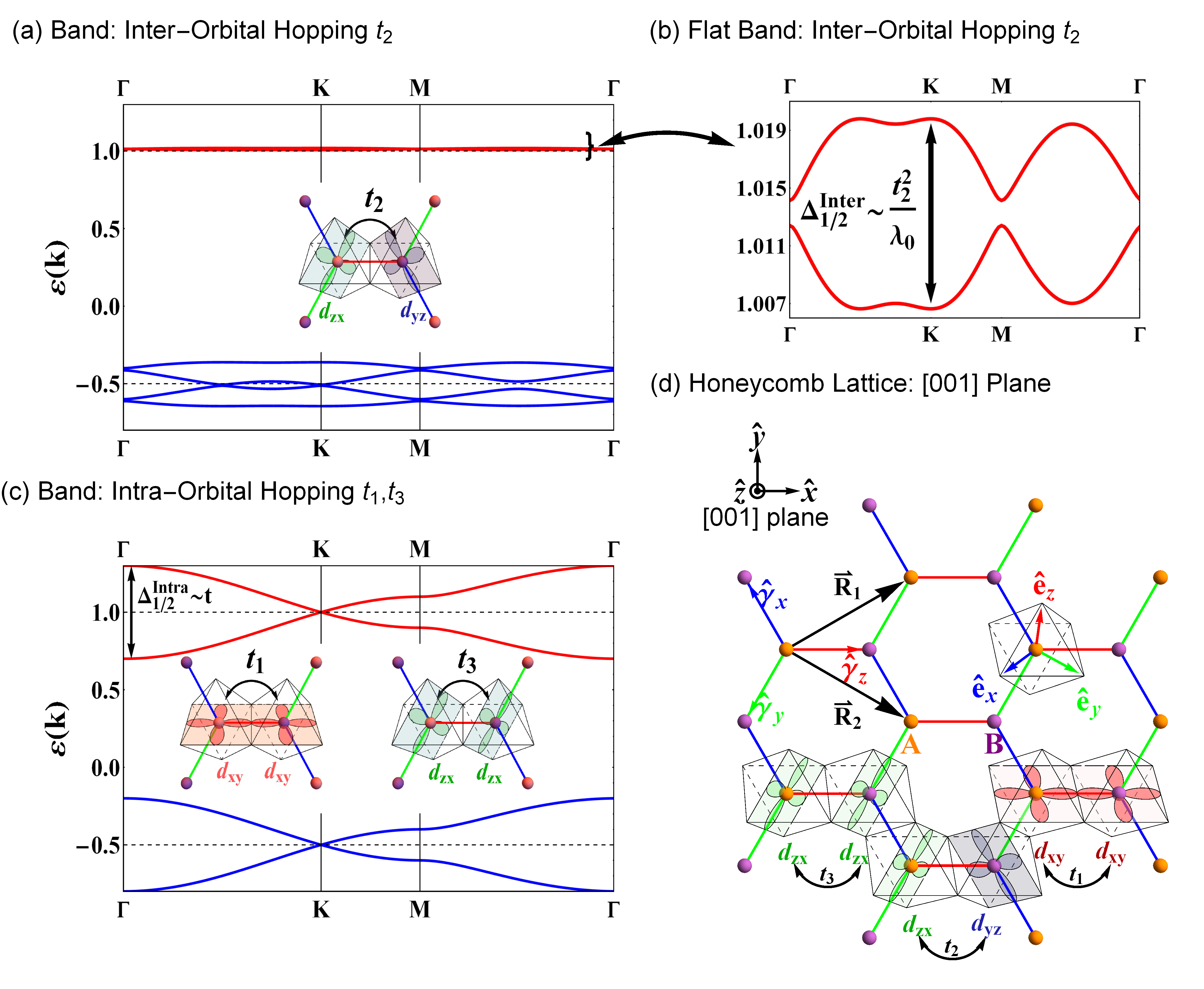}
    \caption{Band structures with spin orbit coupling $\lambda_0$ on a honeycomb lattice: (a) Bands $\varepsilon(\bm{k})$, in units of $\lambda_0$, for dominant inter-orbital hopping $t_2= 0.1 \lambda_0$ and intra-orbital hopping $t_1=t_3=0 $. The bandwidth of upper $j=1/2$ multiplet $\Delta_{1/2}^{\text{inter}}\sim t^2/\lambda_0\sim 0.01\lambda_0$.  (b) Inset highlighting the flat bands in upper $j=1/2$ manifold of (a).  (c) Bands of intra-orbital dominated hopping  $t_1=t_3=0.1 \lambda_0$ and $t_2=0 $. The bandwidth of upper $j=1/2$ multiplet is $\Delta_{1/2}^{\text{intra}}\sim t_1 \sim 0.1 \lambda_0$.  (d) Honeycomb lattice in [001] plane showing orthogonal Cartesian axes $(\hat{\bm{x}},\hat{\bm{y}},\hat{\bm{z}})$, bond axes $(\hat{\bm{\gamma}}_x,\hat{\bm{\gamma}}_y,\hat{\bm{\gamma}}_z)$, and orbital axes $(\hat{\bm{e}}_x,\hat{\bm{e}}_y,\hat{\bm{e}}_z)$ (see Appendix \ref{AppA}), and intersite intra-orbital and intersite inter-orbital processes.}
    \label{fig1_lattice_axes}
\end{figure*}

\medskip
\noindent{\em{Model:}}
The octahedral symmetry around a transition metal ion splits the $d$-orbitals into a 3-fold $t_{2g}$ sector and a 2-fold degenerate $e_g$ sector separated by the crystal field $\Delta_{CF}$ energy. In the low energy $t_{2g}$ sector, the general tight binding Hamiltonian on a honeycomb lattice is given by:

\begin{align}
    \hat{H}&=\sum_{<ij>,\alpha\beta,\sigma}t^{\alpha\beta}_{ij}\hat{d}^\dagger_{\alpha\sigma}(\bm{r}_i)\hat{d}_{\beta\sigma}(\bm{r}_j) \nonumber \\ &+\sum_{i,\alpha\beta,\sigma\sigma'} \lambda^{\alpha\beta}_{\sigma\sigma'}\hat{d}^\dagger_{\alpha\sigma}(\bm{r}_i)\hat{d}_{\beta\sigma'}(\bm{r}_i)
    \label{ham}
\end{align}
\noindent
where $\hat{d}^\dagger_{\alpha\sigma}(\bm{r}_i)$ and $\hat{d}_{\alpha\sigma}(\bm{r}_i)$ are the fermionic annihilation and creation operators at site $\bm{r}_i$ for orbital $\alpha$ and spin $\sigma$.

The first term in Eq.~\ref{ham} describes the hopping between $d$-orbitals at adjacent sites $\langle i,j\rangle$ on the honeycomb lattice and can be written in terms of matrices $T_{\gamma(\bm{r}_i,\bm{r}_j)}$ where $\gamma(\bm{r}_i,\bm{r}_j)$ takes three values $\gamma(\bm{r}_i,\bm{r}_j)=x,y,z$ corresponding to the vector $\bm{v}_s$ that connects the sites $\bm{r}_i$ and $\bm{r}_j$ (see Appendix \ref{AppA}). We write the symmetry allowed components of these matrices $t_{ij}^{\alpha\beta}\delta_{\sigma\sigma'}=T_{\gamma(\bm{r}_i,\bm{r}_j)}^{\alpha\beta}\delta_{\sigma\sigma'}=\bra{d_{\alpha\sigma}(\bm{r}_i)}\hat{t}\ket{d_{\beta\sigma'}(\bm{r}_j)}$ in the basis of orbitals $\{\ket{d_{yz}(\bm{r}_i)},\ket{d_{zx}(\bm{r}_i)},\ket{d_{xy}(\bm{r}_i)}\}$ as

\begin{equation}
    T_x=\left[
    \begin{array}{ccc}
    t_3 & 0 & 0 \\
    0 & t_1 & t_2 \\
    0 & t_2 & t_1
    \end{array}\right],
     T_y=\left[
    \begin{array}{ccc}
    t_1 & 0 & t_2 \\
    0 & t_3 & 0 \\
    t_2 & 0 & t_1
    \end{array}\right],
     T_z=\left[
    \begin{array}{ccc}
    t_1 & t_2 & 0 \\
    t_2 & t_1 & 0 \\
    0 & 0 & t_3
    \end{array}\right]
    \label{Tcoup}
\end{equation}

\noindent
Symmetry constrains the nearest neighbor hopping matrices to have three independent parameters $t_1$, $t_2$, and $t_3$.  The intersite inter-orbital coupling is described by $t_2$, while $t_1$ and $t_3$ describe the intersite intra-orbital couplings.

The second term in Eq.~\ref{ham} describes the onsite spin-orbit interaction $\lambda^{\alpha\beta}_{\sigma\sigma'}\delta_{ij}=\bra{d_{\alpha\sigma}(\bm{r}_i)}\hat{\lambda}\ket{d_{\beta\sigma'}(\bm{r}_j)}$ between the $d$-orbitals in the $t_{2g}$ sector: $\lambda=\dfrac{2\lambda_0}{\hbar^2}\sum_i L_{\hat{\bm{e}}_i}\otimes S_{\hat{\bm{e}}_i}$ ~(see Appendix \ref{AppA}).
Here $S_{\hat{\bm{e}}_i}=(\hbar/2)\sigma_{\hat{\bm{e}}_i}$ and $\sigma_{\hat{\bm{e}}_i}$ is the Pauli matrix in the $\hat{\bm{e}}_i$ direction. In the absence of $\hat{t}$, the spin-orbit interactions splits the $t_{2g}$ sector into states of total angular momentum $j=1/2$ and $j=3/2$. 

\medskip

\noindent {\em {Emergent Flat Bands:}}
By using the discrete translation symmetry of the system, the Hamiltonian in Eq.~\ref{ham} yields a $12\times12$ matrix [$(2j+1)\times 2\ {\rm atoms }$], the Bloch Hamiltonian, $\hat{H}(\bm{k})=e^{-i\hat{\bm{r}}\cdot\bm{k}}\hat{H}e^{i\hat{\bm{r}}\cdot\bm{k}}$ at each ${\bf k}$-point. The eigenvalues of the Hamiltonian
must at least be two-fold degenerate for all crystal momenta $\bm{k}$ due to Kramers degeneracy. In the absence of $\hat{t}$ the bands separate into two manifolds: a manifold with total angular momentum $j=1/2$ containing four bands and a manifold with total angular momentum $j=3/2$ containing eight bands. 

In the purely inter-orbital hopping limit, intra-orbital interactions are turned off ($t_1=t_3=0$) and the orbital interactions are dominated by the inter-orbital coupling $t_2$.  The $j=3/2$ multiplet is split into four two-fold degenerate bands, while the $j=1/2$ multiplet consists of two two-fold degenerate bands with almost no dispersion (see Fig.~ \ref{fig1_lattice_axes} (a)).  The intraband kinetic energy is quenched due to an inter-orbital-spin-orbit frustration caused by the large onsite spin-orbit interaction $\lambda$ and nonzero $t_2$.  This is in striking contrast to the purely intra-orbital ($t_2=0$) limit where bands are seen to disperse in both the $j=1/2$ and $j=3/2$ multiplet (see Fig.~ \ref{fig1_lattice_axes} (c) and Appendix \ref{AppC}).

\medskip


To see the effect of nonzero $\hat{t}$ we use degenerate perturbation theory to calculate the energy spitting of the spin-orbit split bands in orders of the kinetic hopping coefficients $t_1$, $t_2$, and $t_3$.  Remarkably for purely inter-orbital hopping the first order correction to the energy bands of the $j=1/2$ manifold vanishes.  This is due to the orbital character of the $j=1/2$ manifold and the orbital character of the purely inter-orbital hopping matrices.  The action of these matrices on the $j=1/2$ manifold transforms a $j=1/2$ state into a mixture of $j=3/2$ states.  These states are all orthogonal to the $j=1/2$ manifold as they have different eigenvalues with respect to $\hat{\lambda}$.  In order to find the band splitting we must then go to second order perturbation theory leading to an effective bandwidth in the purely inter-orbital hopping limit $\Delta^{\text{inter}}_{1/2}\sim (t_2)^2/\lambda$.  This is in contrast to the band splitting of the $j=3/2$ manifold of states and of either multiplet in the purely intra-orbital hopping limit where the first order correction to the spin-orbit split flat bands is nonzero.  In these cases the action of the kinetic hopping, $\hat{t}$, transforms a state with eigenvalue $j=1/2$ or $j=3/2$ into a linear combination of $j=1/2$ and $j=3/2$ states, leading to a bandwidth $\Delta\sim t$.

These results are most easily determined by writing the Bloch Hamiltonian in the basis of total angular momentum eigenstates $j=1/2,3/2$.  In this basis the Hamiltonian takes the form

\begin{widetext}
\begin{equation}
\hat{H}(\bm{k})=\left(
    \begin{array}{cccc}
      \lambda_0\mathds{1}_{4\times4} +\bm{\mathcal{M}}^{1/2}_k(t_1,t_3)   & \bm{G}_k(t_1,t_2,t_3) \\
      \bm{G}_k^\dagger(t_1,t_2,t_3)  &  -\dfrac{\lambda_0}{2}\mathds{1}_{8\times8}+\bm{\mathcal{M}}^{3/2}_k(t_1,t_2,t_3) 
    \end{array}\right)
\end{equation}
\end{widetext}

\noindent
where $\bm{\mathcal{M}}^{1/2}_k(t_1,t_3)$ is a $4\times 4$ matrix, $\bm{\mathcal{M}}^{3/2}_k(t_1,t_2,t_3)$ is a $8 \times 8$ matrix, and $\bm{G}_k(t_1,t_2,t_3)$ is a $4\times 8$ matrix that are all linear functions of the kinetic couplings $t_i$.  We see that $\bm{\mathcal{M}}^{1/2}_k(t_1,t_3)$ is independent of the interorbital kinetic coupling $t_2$ and thus vanishes in the purely interorbital hopping limit $t_1=t_3=0$, while the other matrices remain non-zero.  This forces the linear in $t$ perturbations to the eigenvalues of the $j=1/2$ manifold in the upper-left $4\times 4$ block of $\hat{H}(\bm{k})$ to vanish, while in this limit the lower-right $8\times 8$ block $\bm{\mathcal{M}}_{3/2}(t_1,t_2,t_3)$ is non-vanishing and leads to the linear in $t$ perturbation to the eigenvalues of the $j=3/2$ sector to be nonzero.  In contrast for the purely intra-orbital hopping limit the matrices $\bm{\mathcal{M}}^{1/2}_k(t_1,t_3)$, $\bm{\mathcal{M}}^{3/2}_k(t_1,t_2,t_3)$, and $\bm{G}_k(t_1,t_2,t_3)$ are all non-zero leading to nonzero first order perturbations to the spin-orbit split flat bands (for details see Appendix \ref{AppC}).

In the presence of nonzero intra-orbital hopping the first order kinetic correction to the spin-orbit split bands is of order $t_1$ and $t_3$.  
In the limit $t_1,t_3<<(t_2)^2/\lambda$ the dispersion of the eigenenergies of the full Bloch Hamiltonian is proportional to the inter-orbital hopping $t_2$ and inversely proportional to the onsite spin-orbit scale $\lambda$, and leads to bandwidths in the $J=1/2$ multiplet of states $\Delta_{1/2}\sim (t_2)^2/\lambda$ similar to the completely frustrated limit ($t_1=t_3=0$).  While in the limit $(t_2)^2/\lambda\lesssim t_1,t_3$ the dominant intra-orbital dynamics diminishes orbital frustration and leads to bandwidths in the $J=1/2$ multiplet $\Delta_{1/2}\sim t_1,t_3$ (see Appendix \ref{AppC}).


The mechanism described above leading to the flat bands in the $J=1/2$ multiplet in the strong inter-orbital hopping limit is just one example of {\it spin-orbit assisted inter-orbital frustration}.  Engineering the multi-orbital onsite potentials and synthetic multi-dimensional inter-site gauge fields that couple onsite degrees of freedom to neighboring sites on the lattice in highly tunable platforms, like cold atoms systems, can provide a framework for exploring orbital frustration and flat bands in different lattice systems with varying onsite and inter-site multi-dimensional spin-orbit potentials \cite{goldman2014light,lin2016synthetic,celi2014synthetic,jimenez2015tunable,cooper2019topological}.

\medskip
\noindent{\em{Quantum Band Geometry and Symmetry Breaking:}}
The quantum geometry of a band structure is determined by a band's Berry potential $\bm{A}_n(\bm{k})=i\bra{u_n(\bm{k})}\bm{\nabla}_k\ket{u_n(\bm{k})} $ and curvature $\bm{\Omega}_n(\bm{k})=\nabla_k\times\bm{A}_n(\bm{k})$,
whose integral across the 2D Brillouin zone dictates the winding of the phase of the Bloch wavefunction: an integer topological invariant called the Chern number. In the purely atomic limit, whereby the hopping of electrons across lattice sites is strictly forbidden, the Berry curvature is trivially zero.  Here we show that in the presence of strong inter-site inter-orbital interactions, $t_2$, the orbital structure  of Bloch eigenstates becomes strongly mixed across the Brillouin zone leading to unique Berry curvature density in the presence of time reversal and inversion breaking perturbations (see Fig~\ref{fig:berrycurve_BrokenT}(B) in Appendix \ref{AppD}).  

The Berry curvature is a pseudo-vector such that for inversion symmetric systems $\bm{\Omega}^k_n(-\bm{k})=\bm{\Omega}^k_n(\bm{k})$ and for time reversal symmetric systems $\bm{\Omega}^k_n(-\bm{k})=-\bm{\Omega}^k_n(\bm{k})$.  In the models described above time reversal and inversion symmetry constrain the Berry curvature to vanish in every band $n$ and at all crystal momentum $\bm{k}$, so $\bm{\Omega}_n(\bm{k})=0 \  {\forall \bm{k}}, {\forall n}$.

In order to observe non-zero Berry curvature we perturb our model system with a time reversal breaking perturbation that couples directly to the electronic spin or with an inversion breaking onsite sublattice potential

\begin{align}
    \hat{H}'_{\mathcal{T}}&=\sum_{i,\sigma\sigma',\alpha} \mu_B\bm{B}_0\cdot\bm{\sigma}_{\sigma\sigma'} \hat{d}^\dagger_{\alpha\sigma}(\bm{r}_i)\hat{d}_{\alpha\sigma'}(\bm{r}_i) \\
    \hat{H}'_{\mathcal{I}}&=\sum_{i,\sigma,\alpha} m_0(\bm{r}_i) \hat{d}^\dagger_{\alpha\sigma}(\bm{r}_i)\hat{d}_{\alpha\sigma}(\bm{r}_i)
\end{align}

\noindent
where $m_0(\bm{r}_i)=+1(-1)$ if site $\bm{r}_i$ is in the $A(B)$ sub lattice.  Below we show the band structure and Berry curvature for the second lowest band in the $j=1/2$ multiplet in the intra-orbital and inter-orbital hopping dominated limits.  

In Fig.~\ref{fig:berrycurve_BrokenT}(A) time reversal is broken and bands in the $j=1/2$ manifold are split in both the purely intra-orbital and purely inter-orbital hopping limits.  In the intra-orbital limit internal orbital symmetry allows bands to cross near the $\bm{K}$ and $\bm{K}'$ points in the Brillouin zone.  In this case the Berry curvature vanishes everywhere except at the points where bands cross and the curvature becomes ill defined.  In the inter-orbital dominated limit, peaks in the Berry curvature occur near the $\bm{M}$ points in the Brillouin zone where the band gap between the lowest and second lowest bands of the $j=1/2$ multiplet becomes narrow. 

Fig.~\ref{fig:berrycurve_BrokenT}(B) shows results for systems with broken inversion symmetry.  In the intra-orbital limit, internal orbital symmetry allows bands in the $j=1/2$ manifold to remain two-fold degenerate at every point in the Brillouin zone (see Appendix \ref{AppD}).  We see that the Berry curvature is localized at the $\bm{K}$ and $\bm{K}'$ points where the band gap between the lower and upper bands of the $j=1/2$ multiplet narrows.  In the inversion broken inter-orbital hopping dominated limit, the narrow bands in the $j=1/2$ multiplet are split everywhere except at the $\bm{\Gamma}$ and $\bm{M}$ points where the bands come together and form two two-fold degenerate points and the Berry curvature is ill defined.  Away from these points the Berry curvature disperses in a three-fold symmetric pattern with its saddle points along high symmetry lines connecting the $\bm{\Gamma}$ and $\bm{M}$ points in the Brillouin zone.

\medskip

\noindent {\em{Optical Signatures:}}
The topological and geometric character of the band structures of the purely intra-orbital hopping and inter-orbital hopping lmits can be distinguished in the system's response to a frequency-dependent electric field.  

The optical conductivity has two contributions:  The intraband contribution $\sigma_F(\omega)$ from electromagnetic field induced transitions on the Fermi surface and the interband contribution $\sigma_I(\omega)$ from transitions between occupied and unoccupied Bloch states with the same crystal momentum $\bm{k}$.  Here we focus on systems where there are ten electrons per unit cell such that the $j=3/2$ multiplet is completely filled and the $j=1/2$ multiplet is half filled.  

In the intra-orbital dominated coupling limit the system is a semimetal whose Fermi surface  contains only two crystal momenta, the $\bm{K}$ and $\bm{K}'$ points, in the Brilloiun zone.  At these points in the band structure the energy states of the $j=1/2$ multiplet become degenerate resulting in two four-fold degenerate points whose low energy effective theory resembles that of two half-filled Dirac cones.  In the inter-orbital dominated limit the system is insulating and the band gap minima move away from the $\bm{K}$ and $\bm{K}'$ points to the $\bm{\Gamma}$ and $\bm{M}$ points.  In both cases the Fermi surface is of vanishing measure leading to negligible intraband contributions to the conductivity at zero temperature.

Symmetry constrains the interband contribution to the conductivity 
$\sigma(\omega)$ (see Appendix \ref{AppE}).  Fig.~ \ref{fig3_conductivity} shows the absorption, the real part of $\sigma(\omega)$, in the extreme intra-orbital and inter-orbital dominated limits for the chemical potential in the center of the $j=1/2$ multiplet.  In the intra-orbital dominated coupling limit the low frequency behavior of $\Re{(\sigma(\omega))}$ is dominated by optical transitions between the lower bands and the upper bands of the $j=1/2$ multiplet near Bloch momenta surrounding the Dirac cones at the $\bm{K}$ and $\bm{K}'$ points in the Brilloiun zone.  This feature saturates at $\hbar\omega\sim 3 t_1=\Delta^{\text{intra}}_{1/2}$ where transitions between Bloch electrons at the band minima and maxima occur.  The absorption vanishes for $\Delta_{1/2}^{\text{intra}}<\hbar\omega<\lambda_0-\Delta^{\text{intra}}_{3/2}/2$ due to the large onsite spin-orbit coupling that splits the $j=1/2$ and $j=3/2$ multiplet by a large energy gap.  For $\hbar\omega>\lambda_0-\Delta^{\text{intra}}_{3/2}/2$ the conductivity vanishes due to a fine tuned extra orbital symmetry inherited from the absence of inter-orbital hopping elements in the extreme intra-orbital dominated coupling limit.  The extra symmetry constrains the interband matrix elements of the velocity operator $\hat{\bm{v}}(\bm{k})$ to vanish between states in different multiplet sectors, forcing the conductivity to vanish at high frequencies.

In the inter-orbital dominated limit the low frequency absorption $\Re{(\sigma(\omega))}$ occurs in a narrow frequency range where light can excite particle-hole pairs within the flat bands of the $j=1/2$ multiplet.  While at higher frequencies $\hbar\omega>\lambda_0-\Delta^{\text{inter}}_{3/2}/2$ the conductivity is finite across a large range of frequencies arising from the joint density of states and the interband matrix elements of the velocity operator between the $j=1/2$ and $j=3/2$ multiplet.

\begin{figure*}[!htb]
    \centering
    \includegraphics[width=0.99\textwidth]{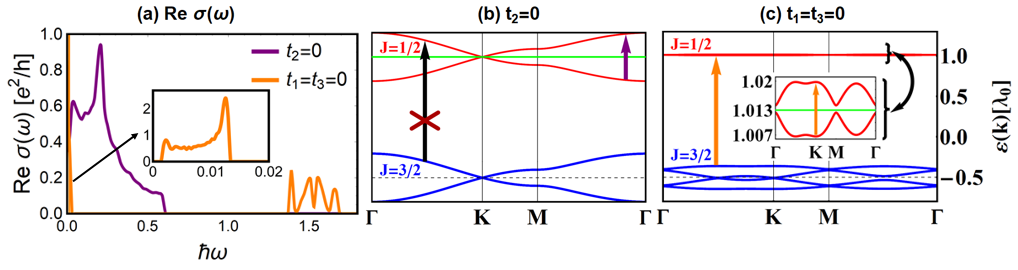}
    \caption{(a) $\Re(\sigma(\omega))$ for system in the purely inter-orbital (orange) $t_2=0.1\lambda_0; t_1=t_3=0$ and purely intra-orbital (purple) $t_1,t_3=0.1\lambda_0; t_2=0$ hopping limits, with the chemical potential (shown in green in (b)) set in the middle of the $j=1/2$ multiplet
    (with $\hbar\omega$ in units of $\lambda_0$).  Selection rules for purely intra-orbital hopping limit, and (c) for purely inter-orbital hopping limit.  Allowed transitions are depicted by colored arrows whose contribution to the conductivity are shown in (a).}
    \label{fig3_conductivity}
\end{figure*}

\medskip
\noindent{\em{Conclusion:}}
We have demonstrated a new avenue for achieving flat bands with a flatness parameter $\mathcal{F}\sim (t_2/\lambda)^2$ in materials with strong onsite spin-orbit interactions $\lambda$ designed to have dominant inter-orbital hopping $t_2$ compared to intra-orbital hopping. 
Flat band generation from electronic interference pathways usually relies on large and complicated unit cell structure \cite{ochi2015robust,arita2002gate,shima1993electronic,liu2013flat}.  Here instead we exploit a particular orbital structure on the honeycomb lattice whose large spin-orbit interaction leads to frustration in the $j=1/2$ manifold of states.

Density functional theory calculations on 4d and 5d transition metal compounds suggest that parameter regimes where $t_2$ dominates over $t_1$ are indeed viable~\cite{kim2016crystal}. Materials based on 5d transition metal ions with larger $\lambda$ and smaller $U$ compared to the 4d transition metal ions should be a particularly attractive design platform for flat band metals that we have focused on here.

In general, {\it spin-orbit assisted inter-orbital frustration} provides a scheme for achieving flat bands in materials with a wide variety of different multi-orbital structures on lattices in any dimension. Gating to control the chemical potential and strain to tune the hopping amplitudes can be used to engineer the orbital frustration needed to generate flat bands.  In addition to quantum materials, cold atom systems have the ability to pattern inter-site lattice interactions in the form of synthetic gauge potentials between multi-orbital unit cells that provide a synthetic dimension thereby creating a definitive platform for engineering orbitally frustrated systems with emergent flat bands in systems with a variety of lattice and multiplet on-site structures. 

We expect these designer flatband materials to provide a link between frustrated metals and frustrated magnetic insulators opening a doorway for the search and discovery of topologically ordered states. 
While the emergence of ferromagnetism in specific flatband systems has been explored  \cite{mielke1993ferromagnetism,tasaki1998nagaoka,lieb1989two,shima1993electronic,tasaki1992ferromagnetism,mielke1991ferromagnetic}, the nature of magnetic interactions in the model described here provides a richer platform.

Controlling orbital frustration on a lattice can lead to material design and discovery of systems on the edge of an instability toward unique correlated and topological many-body phases where the orbital structure of the ground sate plays an important role in the electronic correlations of the system. Rich phenomenology for the proposed flatband metals to show fractionalization of the electronic spectral function, non-obeyance of the Luttinger volume (Luttinger breaking), and quantized anomalous and topological Hall effects need to be further explored.

\medskip

\noindent {\em Acknowledgments:} 
W.Z. acknowledges support from the Center for Emergent Materials: an NSF MRSEC under award number DMR-2011876; Z.A. and N.T. acknowledge support from National Science Foundation Grant number ~DMREF-1629382.

\medskip
\noindent

\bibliography{mybib}

\appendix

\section{Model Geometry}
\label{AppA}

In the main text we demonstrate spin-orbit assisted orbital frustration and the emergence of flat bands in a model containing d-orbital hopping on the honeycomb lattice.  The honeycomb lattice can be described by a primitive triangular lattice with two sites $A$ and $B$ per unit cell.  Each $A$ and $B$ site has three nearest neighboring sites.  We orient the honeycomb such that the primitive lattice vectors can be chosen to be $\bm{R}_1=a\sqrt{3}/2\hat{\bm{x}}+a/2\hat{\bm{y}}$ and $\bm{R}_2=a\sqrt{3}/2\hat{\bm{x}}-a/2\hat{\bm{y}}$, where $a$ is the lattice constant.  We denote the three vectors connecting an $A$ site to its nearest neighboring $B$ sites as $\bm{v}_x=-a/2\sqrt{3}\hat{\bm{x}}+a/2\hat{\bm{y}}$, $\bm{v}_y=-a/2\sqrt{3}\hat{\bm{x}}-a/2\hat{\bm{y}}$, and $\bm{v}_z=a/\sqrt{3}\hat{\bm{x}}$.  These vectors can be related to the bond axes

\begin{align}
    \hat{\bm{\gamma}}_x&=-\dfrac{1}{\sqrt{6}}\hat{\bm{x}}+\dfrac{1}{\sqrt{2}}\hat{\bm{y}}+\dfrac{1}{\sqrt{3}}\hat{\bm{z}}\nonumber \\
    \hat{\bm{\gamma}}_y&=-\dfrac{1}{\sqrt{6}}\hat{\bm{x}}-\dfrac{1}{\sqrt{2}}\hat{\bm{y}}+\dfrac{1}{\sqrt{3}}\hat{\bm{z}}\nonumber \\
    \hat{\bm{\gamma}}_z&=\sqrt{\dfrac{2}{3}}\hat{\bm{x}}+\dfrac{1}{\sqrt{3}}\hat{\bm{z}}
\end{align}

\noindent
whose projection onto the plane of the honeycomb is in the direction of nearest neighbor bond vectors $\bm{v}_s$ (see Fig.~\ref{fig1_lattice_axes} (d)). 

The tight binding model described in equation \ref{ham} can be described by the sum of two terms: a kinetic part describing the overlap between nearest neighboring d-orbitals on the honeycomb lattice and a spin-orbit part describing the onsite interaction between the spin and orbital degrees of freedom of the $t_{2g}$ sector.  It is easiest to write these coupling in a maximally symmetric coordinate system whereby the d-orbitals $\ket{d_\alpha(\bm{r}_i)}$ and spin are defined with respect to the orthogonal orbital unit vectors

\begin{align}
    \hat{\bm{e}}_x&=-\dfrac{1}{\sqrt{2}}\hat{\bm{x}}-\dfrac{1}{\sqrt{6}}\hat{\bm{y}}+\dfrac{1}{\sqrt{3}}\hat{\bm{z}}\nonumber \\
    \hat{\bm{e}}_y&=\dfrac{1}{\sqrt{2}}\hat{\bm{x}}-\dfrac{1}{\sqrt{6}}\hat{\bm{y}}+\dfrac{1}{\sqrt{3}}\hat{\bm{z}}\nonumber \\
    \hat{\bm{e}}_z&=\sqrt{\dfrac{2}{3}}\hat{\bm{y}}+\dfrac{1}{\sqrt{3}}\hat{\bm{z}}
\end{align}

\noindent
that are related to the bond axes by a $\pi/2$ counterclockwise rotation (see Fig.~\ref{fig1_lattice_axes} (d)).

In this basis of orbitals $\{\ket{d_{yz}(\bm{r}_i)},\ket{d_{zx}(\bm{r}_i)},\ket{d_{xy}(\bm{r}_i)}\}$ and along these coordinate axes the kinetic hopping $\hat{t}$ can be written in terms of the matrices in equation \ref{Tcoup}, while the orbital angular momentum operators in the onsite spin-orbit interaction $\hat{\lambda}$ can be written as 

\begin{align}
    L_{\hat{\bm{e}}_x}=\hbar\left(
    \begin{array}{ccc}
    0 & 0 & 0 \\
    0 & 0 & -i \\
    0 & i & 0
    \end{array}\right) \nonumber\\
     L_{\hat{\bm{e}}_y}=\hbar\left(
    \begin{array}{ccc}
    0 & 0 & i \\
    0 & 0 & 0 \\
    -i & 0 & 0
    \end{array}\right) \nonumber \\
     L_{\hat{\bm{e}}_z}=\hbar\left(
    \begin{array}{ccc}
    0 & -i & 0 \\
    i & 0 & 0 \\
    0 & 0 & 0
    \end{array}\right)
\end{align}

\section{Electronic Bloch Bands}
\label{AppB}

The Hamiltonian in Eq.~\ref{ham} gives rise to flat bands in the $j=1/2$ multiplet in the purely inter-orbital coupling limit ($t_1=t_3=0$).  This can be verified by investigating the eigenvalues of the Bloch Hamiltonian $\hat{H}(\bm{k})$.  Our tight binding model has a discrete translation symmetry due to the lattice symmetry of the honeycomb such that its energy eigenstates are Bloch modes $\Psi_n(\bm{k})$ with energy eigenvalues $\varepsilon_n(\bm{k})$ both of which can be labeled by their band index $n$ and crystal momentum $\bm{k}$ taking values in the Brillouin zone.  The Bloch Hamiltonian $\hat{H}(\bm{k})=e^{-i\hat{\bm{r}}\cdot\bm{k}}\hat{H}e^{i\hat{\bm{r}}\cdot\bm{k}}$ is a $12\times12$ matrix we write as

\begin{equation}
    \hat{H}(\bm{k})=\sum_{\mu,\alpha} t_{\mu\alpha}(\bm{k})(\Gamma_\mu\otimes \mathcal{D}_\alpha\otimes \mathds{1})+\sum_{i}\lambda_0(\mathds{1}\otimes L_{\hat{\bm{e}}_i}\otimes S_{\hat{\bm{e}}_i})
    \label{blochham}
\end{equation}

\noindent
Where $\Gamma_\mu$ are the two dimensional Pauli matrices describing the $A/B$ sublattice degrees of freedom, $\mathds{1}$ is the two dimensional identity matrix, and $t_{\mu\alpha}(\bm{k})$ the three dimensional matrices $\mathcal{D}_\alpha$ that satisfy

\begin{align}
    \sum_\alpha t_{x\alpha}(\bm{k})\mathcal{D}_\alpha&=\Re\bigg(\sum_ie^{i\bm{k}\cdot\bm{v}_i}T_i\bigg) \nonumber \\
    \sum_\alpha t_{y\alpha}(\bm{k})\mathcal{D}_\alpha&=\Im\bigg(\sum_ie^{i\bm{k}\cdot\bm{v}_i}T_i\bigg) \nonumber \\
    t_{z\alpha}(\bm{k})&=0
\end{align}

\noindent
where $\bm{v}_i$ are the three vectors defined above that connect an $A$ site to its neighboring $B$ sites.

We now investigate the band structure of our model in the purely intra-orbital ($t_2=0$, $t_1=t_3\neq 0$) and inter-orbital ($t_2\neq 0$, $t_1=t_3=0$) kinetic limits.  For general orbital hopping our model satisfies time reversal symmetry ($\mathcal{T}=i\sigma_{\hat{\bm{e}}_y}K$) and inversion symmetry ($\mathcal{I}=\Gamma_x$) endowed by the honeycomb lattice structure ($[\hat{\mathcal{T}},\hat{H}]=[\hat{\mathcal{I}},\hat{H}]=0$).  These symmetries result in a Kramers degeneracy throughout the Brilloiun zone such that bands must be at least two-fold degenerate for all crystal momenta $\bm{k}$.  In the absence of $\hat{t}$, the bands separate into two manifolds: a manifold with total angular momentum $j=1/2$ containing four bands, and a manifold with total angular momentum $j=3/2$ containing eight bands.  This multiplet structure ($4+8$) remains for $\lambda\gg t$ though $j$ is no longer a good quantum number.

Fig.~\ref{fig1_lattice_axes} (a) and (c) shows the band structure along high symmetry directions in the Brilloiun zone for hopping parameters in the extreme intra-orbital and inter-orbital kinetic limits.  In the intra-orbital dominated limit the absence of inter-orbital hopping ($t_2=0$) results in an additional inter-orbital symmetry such that each band in the $j=3/2$ sector is four-fold degenerate throughout the Brilloiun zone, while the $j=1/2$ sector contains two two-fold degenerate bands protected by time reversal and inversion symmetry (see Fig.~\ref{fig1_lattice_axes} (c)).  Band touching points occur at the $\bm{K}=(2\pi/\sqrt{3}a,2\pi/3a)$ and $\bm{K}'=(2\pi/\sqrt{3}a,-2\pi/3a)$ points in the Brilloiun zone.  This results in an eight-fold degenerate double Dirac point for the $j=3/2$ sector and 4-fold degenerate Dirac points for the $j=1/2$ sector.  Note that neither of these topological degeneracies are protected by crystaline symmetries and can be converted to manifolds containing just two-fold degenerate bands through the introduction of other symmetry allowed inter-orbital interactions and are simply an artifact of the finely tuned nature of the purely intra-orbital coupling limit.

In the limit of exterme inter-orbital kinetics, intra-orbital interactions are turned off ($t_1=t_3=0$) and the orbital interactions are dominated by the inter-orbital coupling $t_2$.  The $j=3/2$ multiplet is split into four two-fold degenerate bands, while the $j=1/2$ multiplet consists of two two-fold degenerate bands with almost no dispersion (see Fig.~ \ref{fig1_lattice_axes} (a)).  The intraband kinetic energy is quenched due to an inter-orbital-spin-orbit frustration caused by the large onsite spin-orbit interaction $\lambda_0$ and nonzero $t_2$ that leads to the emergence of flat bands (see {\em {Emergent Flat Bands}} in the main text).

\section{Perturbation Theory Details}
\label{AppC}

In the absence of $\hat{t}$ the Hamiltonian in equation \ref{ham} consists of a band structure of flat band multiplets of total angular momentum $j=1/2$ and $j=3/2$.  Here we calculate the correction to these flat bands in orders of the kinetic couplings $t_1$, $t_2$, and $t_3$ and show the emergent band dispersion in the $j=1/2$ and $j=3/2$ multiplet in the purely inter-orbital hopping ($t_1=t_3=0$) and purely intra-orbital hopping ($t_2=0$) limits.  We find that the first order correction to the degenerate $j=1/2$ multiplet in the purely inter-orbital hopping limit vanishes leading to small band widths of order $(t_2)^2/\lambda$.

To calculate the energy splitting of the degenerate $j=1/2$ and $j=3/2$ multiplets to linear order in $\hat{t}$ we use degenerate perturbation theory and first calculate the components of the matrix

\begin{equation}
   W^j_{\alpha\beta}(\bm{k})= \bra{\psi^j_\alpha}\sum_{\mu,\alpha} t_{\mu\alpha}(\bm{k})(\Gamma_\mu\otimes \mathcal{D}_\alpha\otimes \mathds{1})\ket{\psi^j_\beta}
\end{equation}

\noindent
where $\ket{\psi^j_\alpha}$ are the eigenstates of $\hat{\lambda}$ with total angular momentum $j$.  The eigenvalues of the matrix $W^j_{\alpha\beta}(\bm{k})$ are the first order corrections to the eigenvalues of $\hat{\lambda}$ \cite{griffiths2018introduction}.  For the $j=3/2$ manifold the eigenvalues of $W^{3/2}_{\alpha\beta}(\bm{k})$ are nonzero and complicated functions of $\bm{k}$.

\begin{table*}[t]
  \centering
\begin{tabular}{ |c|c|c|c|c| }
\hline
 {\bf Intra-orbital Dominated Interactions} & \multicolumn{4}{|c|}{Eigenvalues $[t_1]$} \\ 
 \hline
 $\bm{\Gamma}$ & $9/2$ & $3/2$ & $-3/2$ & $-9/2$ \\  
 \hline
 $\bm{K}/\bm{K}'$ & 0 & 0 & 0 & 0  \\
 \hline
  $\bm{M}$ & $3/2$ & $1/2$ & $-1/2$ & $-3/2$ \\
  \hline
  \hline
 {\bf Inter-orbital Dominated Interactions} & \multicolumn{4}{|c|}{Eigenvalues $[t_2]$} \\ 
 \hline
 $\bm{\Gamma}$ & $\sqrt{1+\sqrt{7/16}}$ & $\sqrt{1-\sqrt{7/16}}$ & $-\sqrt{1-\sqrt{7/16}}$ & $-\sqrt{1+\sqrt{7/16}}$ \\  
 \hline
 $\bm{K}/\bm{K}'$ & $\sqrt{2-\sqrt{15/16}}$ & 0 & 0 & $-\sqrt{2-\sqrt{15/16}}$  \\
 \hline
  $\bm{M}$ &  $\sqrt{1+\sqrt{7/16}}$ & $\sqrt{1-\sqrt{7/16}}$ & $-\sqrt{1-\sqrt{7/16}}$ & $-\sqrt{1+\sqrt{7/16}}$ \\
  \hline
  \end{tabular}
   \caption{Table showing eigenvalues of $W^{3/2}_{\alpha\beta}(\bm{k})$ at high symmetry points in the Brillouin zone in the intra-orbital and inter-orbital dominated kinetic limits.  Eigenvalues are given in units of the nonzero kinetic coupling $t$.  Each eigenvalue is two fold degenerate due to time reversal and inversion symmetry.}
     \label{Wvals}
\end{table*}

Table \ref{Wvals} lists eigenvalues of $W^{3/2}_{\alpha\beta}(\bm{k})$ at some high symmetry momenta in the Brilloiun zone for the intra-orbital and inter-orbital dominated kinetic limits. Each eigenvalue is doubly degenerate because of Kramers theorem and the presence of time reversal and inversion symmetries.  We also note the vanishing of the first order correction to the band energies at the $\bm{K}$ and $\bm{K}'$ momenta in the intra-orbital dominated coupling limit.  This is reflected in the full band structure by the presence of the eight-fold degenerate double Dirac point contained in the $j=3/2$ manifold as plotted in Fig. \ref{fig1_lattice_axes} (c).  In order to maintain this multifold degeneracy higher order corrections to the band dispersion a these points must vanish.  Lastly we note that the first order correction to the band energies predicts a bandwidth $\Delta_{3/2}$ of the $j=3/2$ manifold that is of order the hopping strength ($\Delta\sim t$) for both the intra-orbital and inter-orbital dominated kinetic limits.

This is in striking contrast to the $j=1/2$ manifold of states where $W^{1/2}_{\alpha\beta}(\bm{k})$ is nonzero in the intra-orbital dominated coupling limit, but vanishes in the inter-orbital dominated limit.  For general kinetic hopping $t$ its eigenvalues are

\begin{widetext}
\begin{align}
    \delta E^{(1)}_\pm(\bm{k})=\pm\dfrac{2t_1+t_3}{3}\sqrt{3+4\cos\bigg(\dfrac{\sqrt{3}}{2}k_xa\bigg)\cos\bigg(\dfrac{1}{2}k_ya\bigg)+2\cos\bigg(k_ya\bigg)}
\end{align}
\end{widetext}

\noindent
where each eigenvalue $\delta E_\pm(\bm{k})$ is again two fold degenerate because of Kramers theorem.  In both the intra-orbital and inter-orbital dominated coupling limit $\delta E_\pm(\bm{K})=\delta E_\pm(\bm{K}')=0$.  In the intra-orbital dominated limit ($t_2=0$) this results in the presence of Dirac cones at the $\bm{K}$ and $\bm{K}'$ points in the Brillouin zone (see Fig.~ \ref{fig1_lattice_axes} (c)) as further higher order corrections to the band structure also vanish at these high symmetry points.  Remarkably in the inter-orbital dominated kinetic limit, $t_1=t_3=0$, $W^{1/2}_{\alpha\beta}=0$.  This is due to the orbital character of the $j=1/2$ manifold and the orbital character of the purely inter-orbital hopping matrices.  The action of these matrices on the $j=1/2$ manifold transforms a $j=1/2$ state into a mixture of $j=3/2$ states.  These states are all orthogonal to the $j=1/2$ manifold as they have different eigenvalues with respect to $\hat{\lambda}$ forcing $W^{1/2}_{\alpha\beta}(\bm{k})$ to vanish. This demonstrates that the first order correction to the energy eigenvalues vanish in the $j=1/2$ manifold and for the extreme inter-orbital dominated limit.

To calculate any nonzero dispersion in this limit and in the $j=1/2$ manifold we must proceed to second order degenerate perturbation theory.  However, the calculation is simplified in the inter-orbital dominated limit due to $W^{1/2}_{\alpha\beta}(\bm{k})=0$.  The second order correction to the $j=1/2$ manifold of bands is given by

\begin{equation}
    \delta E^{(2)}_\alpha(\bm{k})=\sum_\beta\dfrac{\bra{\psi_\alpha^{1/2}}\hat{t}(\bm{k})\ket{\psi_\beta^{3/2}}\bra{\psi_\beta^{3/2}}\hat{t}(\bm{k})\ket{\psi_\alpha^{1/2}}}{E^{(0)}_{1/2}-E^{(0)}_{3/2}}
\end{equation}

\noindent
where $E^{(0)}_{1/2}$ and $E^{(0)}_{3/2}$ are eigenvalues of $\hat{\lambda}$ for the $j=1/2$ and $j=3/2$ manifold and where

\begin{equation}
\hat{t}(\bm{k})=e^{-i\bm{k}\cdot\hat{\bm{r}}}\hat{t}e^{i\bm{k}\cdot\hat{\bm{r}}}=\sum_{\mu,\alpha} t_{\mu\alpha}(\bm{k})(\Gamma_\mu\otimes \mathcal{D}_\alpha\otimes \mathds{1})
\end{equation}

\noindent
The difference $E^{(0)}_{1/2}-E^{(0)}_{3/2}=3/2\lambda_0$ is of order the spin-orbit coupling $\lambda_0$, while the numerator will be of order $t_2$ squared.  We see that this leads to a bandwidth in the $j=1/2$ manifold and in the inter-orbital dominated limit of $\Delta^{\text{inter}}_{1/2}\sim (t_2)^2/\lambda_0$ whereas the bandwidths in the purely intra-orbital hopping limit $\Delta^{\text{intra}}_{1/2}$ and the bandwidth of the $j=3/2$ manifold in either limit are all of order $t$.

Thus the ingredients that lead to the supression of band dispersion and emergent flat bands in the $j=1/2$ manifold in the inter-orbital dominated limit are the vanishing of the first order perturbative correction to the band energies deriving from the orbital character of the $j=1/2$ manifold of states and the orbital character of a purely inter-orbital coupling between sites, and the presence of a large spin-orbit interaction for which the $j=1/2$ and $j=3/2$ manifolds are separated by a large energy gap and for which $t_2/\lambda_0$ is small.  This ratio sets the scale for the ratio between the expected bandwidth $\Delta\sim t$ and the actual bandwidth of the $j=1/2$ multiplet in the inter-orbital dominated limit $\Delta^{\text{inter}}_{1/2}\sim (t_2)^2/\lambda_0$.

\section{Strong Coupling Limit:}
\label{AppC2}

Upon adding a strong onsite multiorbital Hubbard interaction to Eq.~\ref{ham} of the form

\begin{equation}
\hat{H}_U=\sum_i\bigg(\dfrac{U-3J_H}{2}(\hat{n}_i-1)^2-2J_H\hat{\bm{S}}_i\cdot\hat{\bm{S}}_i-\dfrac{J_H}{2}\hat{\bm{L}}_i\cdot\hat{\bm{L}}_i\bigg)
\end{equation}

\noindent
where $\hat{n}(\bm{r}_i)$, $\hat{\bm{S}}(\bm{r}_i)$, and $\hat{\bm{L}}(\bm{r}_i)$ are the density, spin, and orbital angular momentum operator at site $\bm{r}_i$,  in the limit of strong spin-orbit coupling $\lambda \gg t$ and strong Hubbard interactions an expansion in powers of $t$ leads to an effective $\mathcal{J}-K-\Gamma$ Hamiltonian for a system where half of the $j=1/2$ multiplet is filled, given by

\begin{align}
    \hat{H}_{eff}&=\sum_{<i,j>}\bigg(\mathcal{J}\hat{\bm{J}}(\bm{r}_i)\cdot\hat{\bm{J}}(\bm{r}_j)+K\hat{J}_{\hat{\bm{e}}_{\gamma}}(\bm{r}_i)\hat{J}_{\hat{\bm{e}}_{\gamma}}(\bm{r_j})\nonumber\\ 
    &+ \sum_{\alpha\beta}\Gamma(\hat{J}_{\hat{\bm{e}}_\alpha}(\bm{r}_i)\hat{J}_{\hat{\bm{e}}_\beta}(\bm{r}_j)+\hat{J}_{\hat{\bm{e}}_\beta}(\bm{r}_i)\hat{J}_{\hat{\bm{e}}_\alpha}(\bm{r}_j))\bigg).
    \label{effH}
\end{align}

\noindent
Here $\gamma\rightarrow \gamma(\bm{r}_i,\bm{r}_j)$ is an implicit function of the lattice sites, $\bm{r}_i$ and $\bm{r}_j$, and $\hat{\bm{J}_i}$ is the total angular momentum operator projected onto the $j=1/2$ state space.

\begin{equation}
    \hat{\bm{J}}=\sum_{m=\pm1/2}\ket{1/2,m_j}\bra{1/2,m_j}(\hat{\bm{S}}+\hat{\bm{L}})\ket{1/2,m_j}\bra{1/2,m_j}
\end{equation}

\noindent
The coupling strengths $\mathcal{J}$, $K$, and $\Gamma$ are given by \cite{PhysRevLett.112.077204,rau2014unconventional}

\begin{align}
    \mathcal{J}&=\dfrac{4}{27}\bigg(\dfrac{6t_1(t_1+2t_3)}{U-3J_H}+\dfrac{2(t_1-t_3)^2}{U-J_H}+\dfrac{(2t_1+t_3)^2}{U+2J_H}\bigg) \nonumber\\
    K&=\dfrac{8J_H}{9}\bigg(\dfrac{(t_1-t_3)^2-3t_2^2}{(U-3J_H)(U-J_H)}\bigg) \nonumber \\
    \Gamma&=\dfrac{16J_H}{9}\bigg(\dfrac{t_2(t_1-t_3)}{(U-3J_H)(U-J_H)}\bigg)\  \cdot
\end{align}

\noindent
We see that in the limit of purely intra-orbital hopping ($t_2=0$) and $t_1=t_3$ but non-zero, the effective Hamiltonian is that of a Heisenberg magnetic ($K=\Gamma=0$), while in the limit of purely inter-orbital hopping ($t_2 \neq 0$) and $t_1=t_3=0$ the effective Hamiltonian is purely Kitaev type with $\mathcal{J}=\Gamma=0$.  In this model the inter-orbital coupling $t_2$ acts as an effective {\it spin}-orbit interaction where the orbital degrees of freedom take the role of an effective {\it spin}.  In the limit $t_1=t_3=0$ this effective interaction gives rise to a pseudo-spin frustration that leads to the effective Kitaev type interaction in Eq.~\ref{effH}.

\begin{figure*}[htb!]
    \centering
    \includegraphics[width=0.99\textwidth]{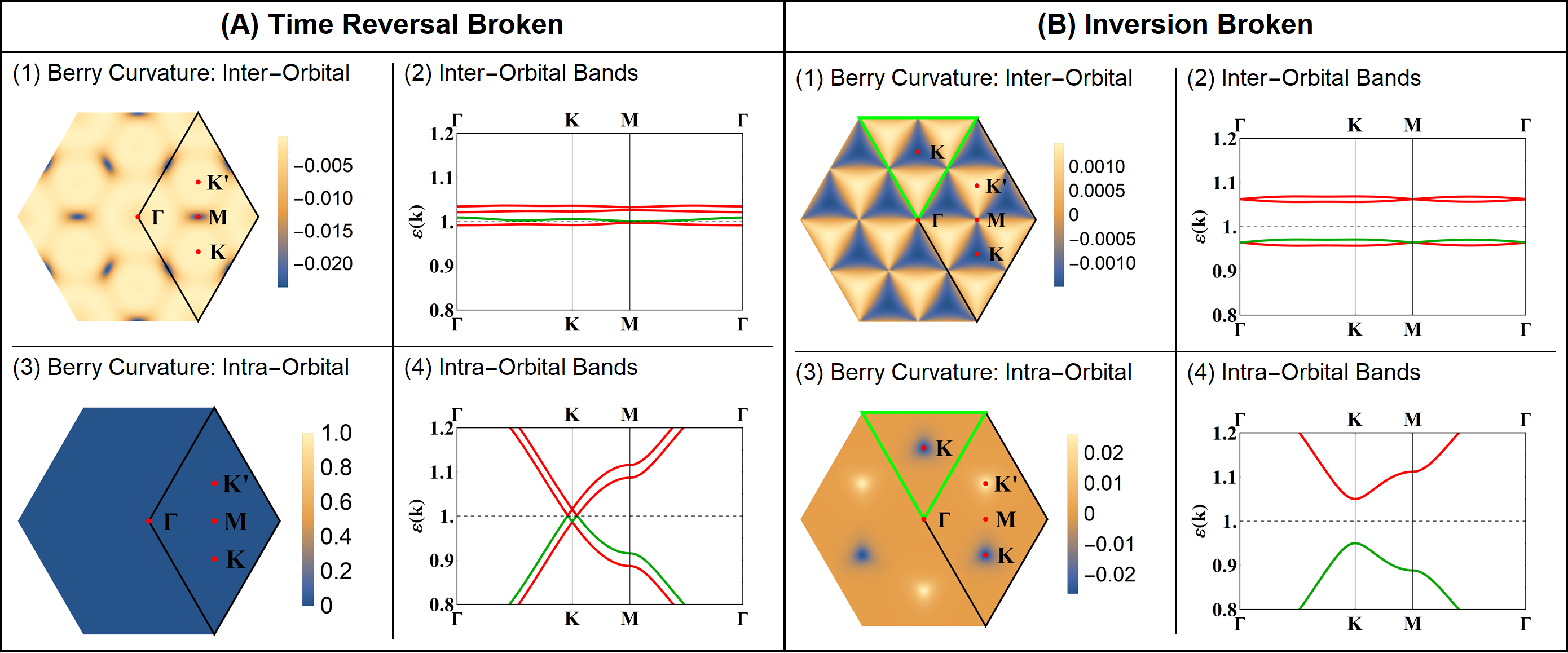}
    \caption{Berry curvature distribution (in units of $a^2$) for [Panels (A)] time reversal broken Hamiltonian $(\mu_B|\mathbf{B}_0|=0.05\lambda_0)$
    and [Panels (B)] for inversion broken Hamiltonian $(|m_0|=0.05\lambda_0)$:
    (1) Berry curvature of {\color{ForestGreen} $10_{\text{th}}$} band; (2) band structure of $9_{\text{th}}$ to $12_{\text{th}}$ bands in purely inter-orbital hopping limit $(t_2=0.1\lambda_0,t_1=t_3=0)$; (3) Berry curvature of {\color{ForestGreen} $10_{\text{th}}$} band; (4) band structure of $9_{\text{th}}$ to $12_{\text{th}}$ bands for purely intra-orbital limit $(t_1=t_3=0.1\lambda_0, t_2=0)$.  The regions outlined in green in panel (B) figures (1) and (3) show the fractionalization of the Berry curvature distribution around the $\bm{K}$ and $\bm{K}'$ points in the Brillouin zone as one interpolates between the purely intra-orbital and purely inter-orbital hopping limits. For full band structures in the presence of both inversion and time reversal symmetry breaking and the Berry curvature distribution and Chern numbers of all 12 bands in the presence of time reversal symmetry breaking and in the purely inter-orbital hopping limit see Fig. \ref{fig_full_bands_TPbroken} and Fig. \ref{fig_chern_number_Tbroken} in Appendix \ref{AppCOND}.} 
    \label{fig:berrycurve_BrokenT}
\end{figure*}

\medskip

\section{Berry Curvature for Degenerate Bands}
\label{AppD}

In order to understand the quantum geometry of the bands that make up the $j=1/2$ multiplet we calculate the momentum space Berry curvature distribution for systems with broken time-reversal and inversion symmetries in both the purely inter-orbital hopping ($t_1=t_3=0)$ and purely intra-orbital hopping limits ($t_2=0$).  In the limit of purely intra-orbital hopping ($t_2=0$) and in the presence of an inversion symmetry breaking mass the $j=1/2$ multiplet contains two two-fold degenerate bands (see Fig.~\ref{fig:berrycurve_BrokenT} (B)(4)). The single band Berry curvature is ill defined in the presence of degenerate bands and thus for this system we plot the Berry phase density for a state with chemical potential filling half the $j=1/2$ multiplet. We first partition the Brillouin zone into small plaquettes and compute the Berry flux through each plaquette

\begin{equation}
    \phi(\bm{k})=-\Im(\text{Ln}(\det(\bm{\mathcal{U}})))
\end{equation}

\noindent
where $\bm{\mathcal{U}}_{nm}=\braket{u_n(\bm{k}_N)|u_m(\bm{k}_0)}$ with $n$ and $m$ taking values over all bands up to the half filled $j=1/2$ multiplet, and where the state $\ket{u_n(\bm{k}_N)}$ is calculated by parallel transport of the state $\ket{u_n(\bm{k}_0)}$ from $\bm{k_0}$ around a plaquette centered at $\bm{k}$ \cite{vanderbilt2018berry}.  The Berry curvature is then found by dividing $\phi(\bm{k})$ by the area of the plaquette.

\section{Hall Conductivity in the Presence of Time-Reversal Symmetry Breaking}
\label{AppCOND}

The quantum geometry of a band structure is determined by a band's Berry curvature, $\bm{\Omega}_n(\bm{k})$,
whose integral across the 2D Brillouin zone dictates the winding of the phase of the Bloch wavefunction: an integer topological invariant called the Chern number.  In the presence of a time-reversal breaking perturbation to the Hamiltonian in equation \ref{ham} both the Berry curvature and Chern number can be nonzero.  Fig. \ref{fig_chern_number_Tbroken} shows the momentum space Berry curvature distributions and Chern numbers for all 12 bands in the presence of time reversal symmetry breaking and in the purely inter-orbital coupling limit  ($t_2\neq 0$,$t_1=t_3=0$).

\section{Optical Conductivity}
\label{AppE}

To understand the optical response of our model to external frequency dependent electric field perturbations we calculate the optical conductivity tensor.  The conductivity tensor $\sigma_{ij}(t,t')\in \mathcal{R}$ such that $\sigma_{ij}(-\omega)=\sigma_{ij}^*(\omega)$.  Furthermore time reversal symmetry constrains the antisymmetric part of the conductivity to vanish $\sigma_{ij}(\omega)=\sigma_{ji}(\omega)$ \cite{onsager1931reciprocal,morimoto2018nonreciprocal}.  Crystalline three-fold rotational symmetry about a lattice point constrains the diagonal parts of the conductivity to be equal ($\sigma(\omega)=\sigma_{xx}(\omega)=\sigma_{yy}(\omega)$) while inplane two-fold rotational symmetry constrains $\sigma_{xy}(\omega)=\sigma_{yx}(\omega)=0$ \cite{utermohlen2020symmetry}.  As described in the main text the intra-band conductivity vanishes at $T=0$ for systems filling half of the $j=1/2$ multiplet, and thus we focus on calculating the real symmetric part of the interband optical conductivity, $\Re{(\sigma_{ij}(\omega))}=\Re(\sigma_I^{ij}(\omega)+\sigma_I^{ji}(\omega))/2$,
that couples to real optical transitions between Bloch states whose energy differs by $\omega$ \cite{sipe1993nonlinear,passos2018nonlinear}.

The real symmetric part of the interband conductivity can be written as

\begin{align}
    \Re(\sigma_{ij}(\omega))&=\dfrac{e^2\pi\hbar}{V}\sum_{nm\bm{k}}\dfrac{f^T_n(\bm{k},\mu)-f^T_m(\bm{k},\mu)}{\varepsilon_n(\bm{k})-\varepsilon_m(\bm{k})} \nonumber \\
     &\times \Re(v^i_{nm}(\bm{k})v^j_{mn}(\bm{k})) \delta(\varepsilon_n(\bm{k})-\varepsilon_m(\bm{k})+\hbar \omega)
\end{align}

\noindent
 Here $V$ is the volume of the system, $e$ is the electric charge, $f^T_n(\bm{k},\mu)$ are the Fermi occupation functions at temperature $T$ and chemical potential $\mu$, and $v_{nm}^i(\bm{k})$ are the interband matrix elements of the velocity operator in the $\hat{\bm{i}}$-direction.

\begin{equation}
    v_{nm}^i(\bm{k})=\dfrac{1}{\hbar}\bra{u_n(\bm{k})}\dfrac{\partial \hat{H}(\bm{k})}{\partial k_i}\ket{u_m(\bm{k})}
\end{equation}

\noindent
Where we have chosen the phase on the periodic part of the Bloch functions $\ket{u_n(\bm{k})}$ such that the full Bloch functions $\ket{\Psi_n(\bm{k})}=e^{i\bm{k}\cdot\hat{\bm{r}}}\ket{u_n(\bm{k})}$ are periodic in the Brilloiun zone $\ket{\Psi_n(\bm{k})}= \ket{\Psi_n(\bm{k}+\bm{G})}$. 

\begin{figure}[!h]
    \centering
    \includegraphics[width=0.49\textwidth]{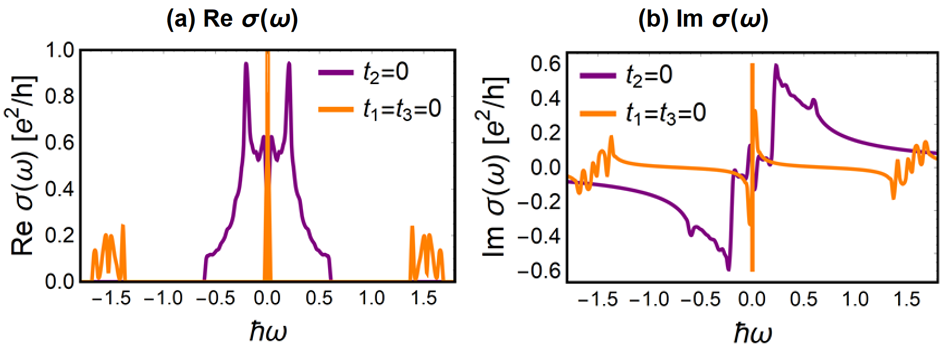}
    \caption{Optical conductivity:  (a) real part and  (b) imaginary part obtained via Kramers-Kronig relations for purely inter-orbital (orange) $t_2=0.1\lambda_0; t_1=t_3=0$ and purely intra-orbital (purple) $t_1,t_3=0.1\lambda_0; t_2=0$ hopping limits.}
    \label{fig5_conductivity_imaginary}
\end{figure}

Fig. \ref{fig3_conductivity} shows the real part of the optical conductivity in the purely inter-orbital and purely intra-orbital coupling limits.  To calculate the imaginary part of $\sigma(\omega)$ we use the Kramers-Kronig relations for complex functions analytic in the upper-half plane of $\omega$: $\Im(\sigma(\omega))=-\dfrac{1}{\pi}\mathcal{P}\int_{-\infty}^{\infty}\dfrac{\Re(\sigma(\omega'))}{\omega'-\omega}$ \cite{kronig1926theory,kramers1927diffusion}.  Fig.~\ref{fig5_conductivity_imaginary}(b) shows the imaginary part of $\sigma(\omega)$ in the two limits of interest for systems with chemical potential where half of the $j=1/2$ multiplet is filled.

\begin{figure*}[!htb]
    \centering
    \includegraphics[width=0.98\textwidth]{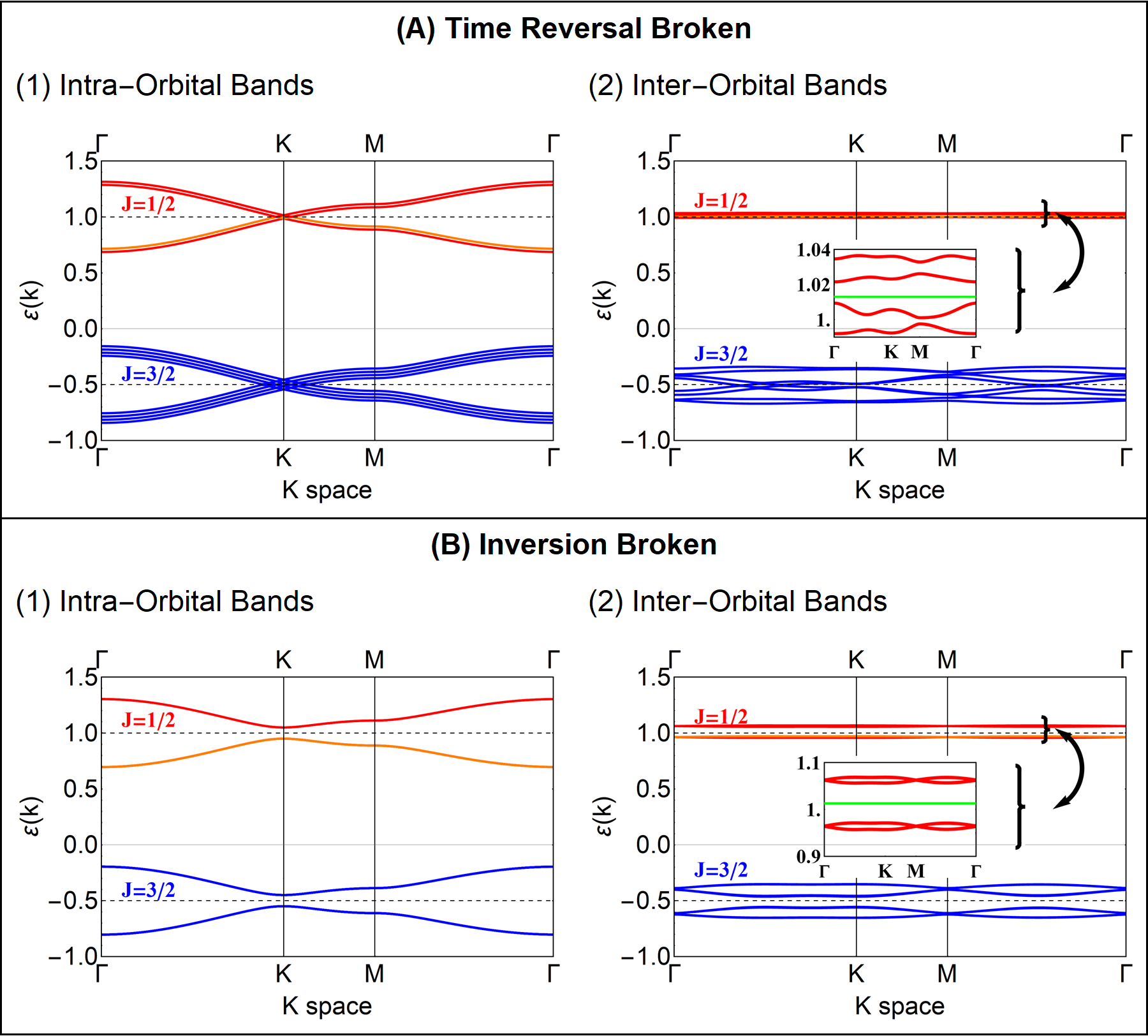}
    \caption{Band structure for model with (A) time reversal symmetry breaking ($\mu_B|\mathbf{B}_0|=0.05\lambda_0$) and (B) inversion symmetry breaking ($|m_0|=0.05\lambda_0$).}
    \label{fig_full_bands_TPbroken}
\end{figure*}

\begin{figure*}[!htb]
    \centering
    \includegraphics[width=0.98\textwidth]{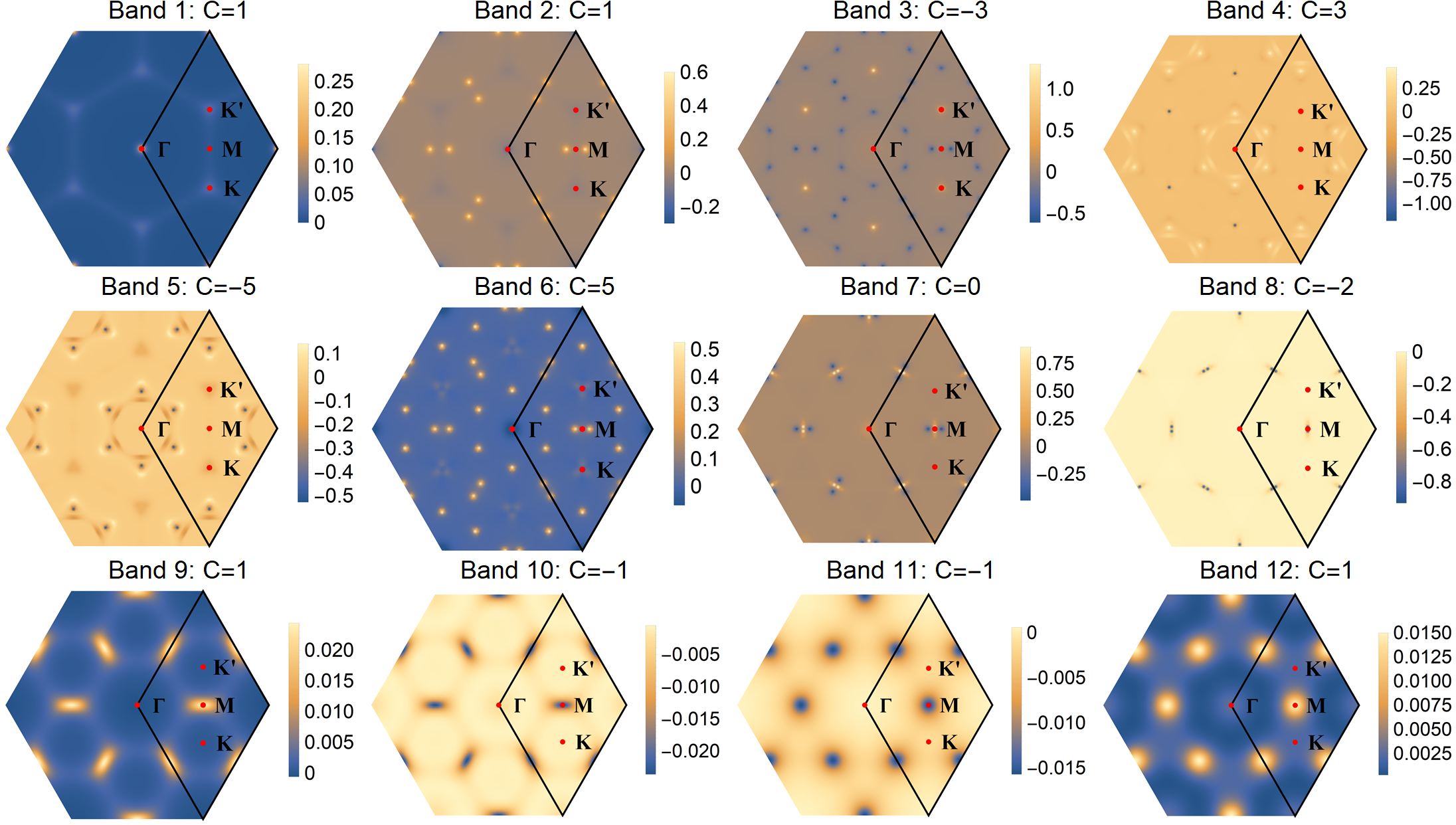}
    \caption{Momentum space Berry curvature distributions and Chern numbers for all 12 bands in the presence of a time reversal breaking perturbation ($\mu_B|\mathbf{B}_0|=0.05\lambda_0$) and in the purely inter-orbital hoping limit ($t_1=t_3=0$).}
    \label{fig_chern_number_Tbroken}
\end{figure*}

\end{document}